\documentclass[10pt,twocolumn]{article}
\usepackage[top=1.9cm, bottom=1.9cm, left=1.8cm, right=1.8cm]{geometry}
\usepackage{times}  %
\usepackage[sort&compress,numbers]{natbib}  %
\usepackage[hyphens]{url}
\usepackage{graphicx}  %
\usepackage[keeplastbox]{flushend}
\usepackage[hyphens]{url}  %
\usepackage{graphicx} %
\usepackage{tikzsymbols}

\newcommand{\descr}[1]{\smallskip\noindent\textbf{#1}}

\usepackage{sectsty}
\sectionfont{\bfseries\Large\raggedright}

\usepackage{url}                                %
\usepackage{array,multirow,graphicx,adjustbox}  %
\usepackage{booktabs}                           %
\usepackage[utf8]{inputenc}
\usepackage[ruled,algosection,noend,linesnumbered]{algorithm2e}
\usepackage{float}
\usepackage{paralist}
\usepackage{amsmath}
\usepackage{amsfonts}
\usepackage{tcolorbox}
\usepackage{xcolor}
\usepackage{csquotes} 
\usepackage{tabularx}
\usepackage{makecell}
\usepackage{pbox}
\usepackage[hang,flushmargin]{footmisc}
\usepackage{flushend}
\usepackage{xspace}
\usepackage{multicol, lipsum}
\usepackage[T1]{fontenc}

\usepackage{xcolor}
\usepackage{booktabs}
\usepackage{graphicx}
\usepackage{paralist}
\usepackage[small,bf]{caption}
\usepackage{subcaption}

\let\oldbibliography\thebibliography
\renewcommand{\thebibliography}[1]{%
  \oldbibliography{#1}%
  \setlength{\itemsep}{2pt}%
}

\usepackage[hang,flushmargin]{footmisc}

\usepackage[compact]{titlesec}
\titlespacing*{\section}{0pt}{*3}{3pt}
\titlespacing*{\subsection}{0pt}{*2}{2pt}
\usepackage{xspace}

\makeatletter
\def\url@leostyle{%
  \@ifundefined{selectfont}{\def\UrlFont{}}%
  {\def\UrlFont{}}%
}
\makeatother
\usepackage[hyphenbreaks]{breakurl}

\usepackage[bookmarks=true, bookmarksnumbered=true, colorlinks=true, linkcolor=linkcol, citecolor=citecol, urlcolor=urlcol, hypertexnames=true]{hyperref}

\definecolor{darkgreen}{RGB}{0, 100, 0}
\definecolor{linkcol}{rgb}{0.3,0,0}
\definecolor{citecol}{rgb}{0.3,0,0}
\definecolor{urlcol}{rgb}{0.3,0,0}

\makeatletter
\def\url@leostyle{%
  \@ifundefined{selectfont}{\def\UrlFont{\small}}%
  {\def\UrlFont{}}%
}
\makeatother
\newif
\ifcomment
\commentfalse
\ifcomment
\newcommand{\sz}[1]{{\bf \textcolor{brown}{SZ: #1}}}
\newcommand{\edc}[1]{{\bf \textcolor{green}{EDC: #1}}}
\newcommand{\gs}[1]{{\bf \textcolor{red}{GS: #1}}}
\newcommand{\ppaudel}[1]{{\bf \textcolor{blue}{PP: #1}}}
\newcommand{\jbnote}[1]{{\bf \textcolor{magenta}{JB: #1}}}
\else
\newcommand{\sz}[1]{}
\newcommand{\edc}[1]{}
\newcommand{\gs}[1]{}
\newcommand{\ppaudel}[1]{}
\newcommand{\jbnote}[1]{}
\fi
\newcommand{\basekeyword}[1]{\textbf{\textit{base keywords}}}
\newcommand{\spanningsubset}[1]{\textbf{\textit{spanning subset}}}
\newcommand{\initialtrainmodel}[1]{\textbf{\textit{initial train model}}}
\newcommand{\initialtrainclaims}[1]{\textbf{\textit{initial train claims}}}
\newcommand{\validationclaims}[1]{\textbf{\textit{reddit validation claims}}}
\newcommand{\twittervalidationclaims}[1]{\textbf{\textit{twitter validation claims}}}
\newcommand{\expandedclaims}[1]{\textbf{\textit{expanded claims}}}
\newcommand{\expandedmodel}[1]{\textbf{\textit{expanded model}}}

\begin{document}
\sloppy

\author{Pujan Paudel$^1$, Jeremy Blackburn$^2$, Emiliano De Cristofaro$^3$, Savvas Zannettou$^4$, and Gianluca Stringhini$^1$\\[0.5ex]
\normalsize{$^1$Boston University, $^2$Binghamton University, $^3$University College London, $^4$Max Planck Institute for Informatics}\\
}
\date{}

\title{\bf Soros, Child Sacrifices, and 5G: Understanding the Spread of Conspiracy Theories on Web Communities}

\maketitle

\begin{abstract}
This paper presents a multi-platform computational pipeline geared to identify social media posts discussing (known) conspiracy theories.
We use 189 conspiracy claims collected by Snopes, and find 66k posts and 277k comments on Reddit, and 379k tweets discussing them.
Then, we study how conspiracies are discussed on different Web communities and which ones are particularly influential in driving the discussion about them.
Our analysis sheds light on how conspiracy theories are discussed and spread online, while highlighting multiple challenges in mitigating them.

\end{abstract}

\section{Introduction}

Conspiracy theories have been a constant presence throughout modern history, %
and the rise of online social networks has significantly broadened their reach.
Users are increasingly exposed to them, from plots to control people through COVID vaccines~\cite{jamison2020not} to claims of mass shootings being false flag incidents~\cite{starbird2017examining} or that child-eating cabals are controlling the United States~\cite{aliapoulios2021gospel}.
These conspiracy theories can have serious adverse effects on society, from fostering polarization to hindering the adoption of public health measures.

As a result, automatically tracking the spread of conspiracy theories is key to understanding their reach/how their narratives evolve and to devising effective mitigations. 
However, developing robust computational systems to track them on social media is challenging.
Conspiratorial discussion changes over time and different Web communities likely discuss them in a markedly different manner, making automated detection systems trained on one community not generalize on another one.
To address this problem, previous work developed tailored techniques that either focus on a single or a handful of conspiracy theories~\cite{samory2018conspiracies,starbird2017examining}, or work on a single online service~\cite{bessi2015science,conti2017s,del2016spreading,tangherlini2020automated,zollo2017debunking}.
However, an effective approach requires tracing any new conspiracy on multiple online services at once. 

In this paper, we present a multi-platform computational pipeline to identify social media messages related to a set of known conspiracy claims.
We start by looking at 189 conspiracy claims debunked by the Snopes fact-checking organization.
To identify social media posts discussing a particular conspiracy theory, we develop an effective Learning-to-Rank technique that can approximate the performance of a human analyst when identifying conspiratorial discussion.
We then collect relevant data from seven subreddits, including the ``usual suspects'' as well as non-conspiracy-oriented subreddits like /r/news, /r/Conservative, and /r/politics.
We also gather data from Twitter.

We show that training our approach on Reddit data produces an accurate detector when tested on Twitter, demonstrating that our system can generalize to multiple online services.
Our approach lets us understand how the same conspiracy theory is discussed on different social networks and how different communities influence each other regarding conspiratorial discussions.

Between 2016 and 2021, we identify 66,274 posts and 288,721 comments on Reddit, and 379,708 tweets on Twitter discussing 189 conspiracy claims.
We then conduct experiments to better understand how conspiracy theories are discussed on different communities by looking at how long they are discussed for, measuring toxicity, and comparing language using word embedding models. 
Finally, we use Hawkes Processes~\cite{hawkes1971spectra} to identify which Web communities are more influential in spreading conspiracy theories (i.e., conspiracy claims posted on them predict those posted elsewhere). %

In summary, we make the following findings:

\begin{itemize}
 \item Conspiracy theories on both Twitter and Reddit are discussed for long periods;  80\% of them for longer than a year.
 \item Reddit comments discussing conspiracy theories show higher levels of toxicity than general discussion on the same communities and Twitter.
 On the other hand, submissions discussing conspiracy theories are the least toxic.
 \item Conspiratorial discussion on /r/Conservative, /r/Worldnews, /r/News, and /r/AskReddit is largely similar, while communities dedicated to conspiracy theories as well as Twitter use different language.
Language in conspiracy-oriented subreddits investigates the details and invites further discussion about them, whereas the four closest communities mostly stick to reporting and referring to the stories.
 \item Different Web communities are influential in discussing different types of conspiracy theories.
   General-purpose subreddits are more influential than those dedicated to conspiracy theories, where users seem to be dissecting their details rather than aiming to make them go viral.
\end{itemize}

Overall, our work provides a solid foundation for computational studies that aim to study conspiracy theories from a large-scale, cross-platform perspective.

\section{Related Work}

In this section, we review previous research analyzing conspiracy theories on different social media platforms.

\descr{Conspiracy theories.}
Conti et al.~\cite{conti2017s} identified conspiracy theories on Facebook using the structural features of the information cascade.
However, their approach yields a relatively low F1 score (65\%).
Tangherlini et al.~\cite{tangherlini2020automated} developed an automated pipeline to discover conspiracy theories and the narrative frameworks surrounding them.
Their work presented generative narrative frameworks on social media, focusing on two popular conspiracy theories: Pizzagate and Bridgegate.

\descr{Facebook-based measurements.}
Zollo et al.~\cite{zollo2015emotional} collected posts from 280K Facebook users on conspiracy-theory and science-related pages.
Discussion on the former are more negative than the latter.
In follow-up work, Zollo et al.~\cite{zollo2017debunking} studied how 54M Facebook users interacted with science-related and conspiracy theory news, reporting the existence of distinct community structures with highly echo-chamber-like behavior between the two communities of users.
These results were somewhat confirmed, from a different perspective, by Del Vicario et al.~\cite{del2016spreading} through the lens of information cascade dynamics.
Bessi et al.~\cite{bessi2015science} studied the consumption pattern of news and conspiracy theories on Facebook, reporting that polarized users contribute more to the diffusion of conspiracy theories.

\descr{Event-based measurements.}
Another line of work studies conspiracy theories on Twitter regarding specific events. %
Starbird~\cite{starbird2017examining} performed a mixed-method analysis on the alternative media ecosystem regarding the spread of conspiracy theories about mass shooting events.
She discussed how alternate media propagate and shape alternative narratives through graph analysis of the URLs being shared on Twitter.
Samory et al.~\cite{samory2018conspiracies} studied four tragic events (i.e., Boston marathon bombing, Sandy Hook shooting, Aurora theater shooting, and the Malaysia Airlines flight MH17 disaster) and 10 years of discussions on the primary conspiracy discussion subreddit, /r/conspiracy.

\descr{Generalized measurements.}
Samory et al.~\cite{samory2018government} developed a scalable method to examine the nature of conspiratorial discussions in online communities.
They analyzed over ten years of discussions in /r/conspiracy by building \textbf{\textit{agent-action-target}} triplets in conspiratorial statements, grouping them into clusters of conspiracies, and identifying themes of conspiracy discussions.
Analyzing the linguistic aspects of conspiracy communities, Klein et al.~\cite{klein2018topic} used topic modeling methods to reveal distinct interests of users within an online conspiracy forum.

Further, Klein et al.~\cite{klein2019pathways} studied the social and linguistic precursors of involvement in conspiracy discussing employing a retrospective case-control study design.
Their analysis reports consistent differences in usage of language between users who eventually join the conspiracy communities and users who do not.
Following a similar theme, Phadke et al.~\cite{phadke2021makes} analyzed the factors contributing to the users joining conspiracy communities by studying longitudinal data from 56 different conspiracy communities on Reddit, finding that mutual interactions with conspiracy communities and marginalization outside the conspiracy communities play the most critical role on Reddit users joining conspiracy communities.
Finally, Phadke et al.~\cite{phadke2021characterizing} used a mixed-methods approach to characterize the social imaginaries in conspiracy communities and the various dimensions of their language.

Our study differs from previous work in two different ways: first, we present a generalized end-to-end pipeline starting from a list of conspiracy claims, officially debunked by Snopes (easily extendable to any other source list of claims), to analyzing the language and influence between communities surrounding the conspiracy.
Second, our work studies discussion of conspiracy theories between different types of communities (conspiracy-oriented and non-conspiracy-oriented subreddits) within a social media platform and across social media platforms (Reddit and Twitter).
 
\section{Methodology and Datasets}
This section presents our dataset and the methodology used to identify and analyze the discussion of conspiracy theories across multiple Web communities.
Our analysis pipeline consists of the following five components:

\begin{enumerate}
\item
 \textit {Claim and social media data collection}: Selecting conspiracy theory claims from Snopes and identifying Web communities to search for conspiratorial discussions.
\item {\em Keyword Identification}: Extracting the keywords from conspiracy claims and learning to rank the keywords for automated extraction over the entire conspiracy claims.
\item
\textit {Filtered Data Extraction}: Extracting relevant posts and tweets from the extracted keywords for further analysis.
\item
\textit {Discussion of Conspiracies}: Analyzing the language of discussion of conspiracy theories within the platforms using word embeddings.
\item
\textit {Influence Estimation}: Assessing the influence Web communities have on each other wrt spreading the conspiracy theories.
\end{enumerate}

\subsection{Claim collection and social media data}
\label{section:datacollection}
\descr{Conspiracy claims.}
We begin by extracting the conspiracy claims for the conspiracy theories listed by the fact-checking organization Snopes.
Snopes does not present their operational definition of a conspiracy theory nor distinguish it from the different claims they investigate on their website.
Following Zannettou et al.~\cite{zannettou2019web}, we use the operational definition of conspiracy theories as ``stories that try to explain a situation or an event by invoking prior closely related stories: without proof, adopting an evidence-based approach relying on leaps of faiths to connect ambiguous actors and events.''
Conspiracy theories are mostly about actions of governments or influential individuals, and range from discussing ``Illuminati'' to false flags on mass crisis scenarios.

Conspiracy theories on Snopes are posted under the section Politics; hence, we can expect a ``bias'' toward political conspiracies.
Every entry on Snopes includes a claim title and a subtitle section.
The former is mostly framed as questions, consisting of the subject of the conspiracy, the actions that invoked the conspiracy theory, and other actors who could be affected by it.
Some examples of claim titles are: ``Did AT\&T have a contract to audit Dominion voting systems,'' and ``No, China is not amassing troops in Canada to invade the US.''
From each article in the conspiracy section, we extract the claim title as the source text.
We discard subtitles, as they often ignore the primary actors of the particular conspiracy claim and are primarily used to provide additional context.
Overall, we collect a total of 189 conspiracy claims in our dataset, spanning from June 2016 to April 2021.

\descr{Reddit.}
Reddit is a social news website and forum, made up of communities where content is socially curated and promoted by members of the community through voting.
We begin our data collection with the primary conspiracy discussion community of Reddit, /r/conspiracy.
We then add other subreddits that are most similar to /r/conspiracy; to do so, we rely on the work by Phadke et al.~\cite{phadke2021makes}, who identify the ancillary communities related to /r/conspiracy by calibrating a conspiracy scale on Reddit communities.
We select the top 10 subreddits from~\cite{phadke2021makes}, which we collectively identify as \emph{conspiracy-oriented communities} in the rest of this paper.
Naturally, conspiracy theories might be discussed on non-conspiracy-oriented subreddits as well.
Hence, we select six subreddits: /r/news, /r/worldnews, /r/democrats, /r/politics, /r/Conservative, and /r/The\_Donald as what we defined as {\em non-conspiracy-oriented communities}.
We do so to understand the nature of discussion, and the influence of these non-conspiratorial communities.

Overall, we use the monthly dumps available from Pushshift~\cite{baumgartner2020pushshift} to extract the Reddit data, from January 2016 to September 2021.
Although conspiracy claims gathered from Snopes only span June 2016--April 2021, we also add additional time windows to let us factor in the conversation about conspiracy theories that might occur before or after Snopes publication. %

\descr{Twitter}.
We collect tweets available through the 1\% Public Streaming API between January 1, 2016, and August 31, 2021.

\descr{Ethical Considerations}.
We only use data published publicly on Web communities and do not interact with users in any way.
As such, this research is not considered human subjects by the IRB at our institution.
Nonetheless, we follow standard ethical guidelines; for example, we make no attempts to de-anonymize users.

\subsection{Keyword Identification}
\label{section:learningtorank}

As a first step, we need to identify Reddit posts and comments as well as tweets that discuss specific conspiracy claims.
To this end, we apply the Learning to Rank (LTR) technique~\cite{cao2007learning} to determine, given a document title (i.e., the Snopes Claim) and a document store (i.e., social media posts), the optimal set of keywords that retrieve the best results from the document store.

To train the LTR model, we begin by building a ground truth dataset for a fraction of the conspiracy claims we collected, manually labeling relevant posts and tweets.
We then develop features to train the LTR model and evaluate it on unseen claims to test the model's effectiveness.
We also compare our LTR model with existing keyword extraction approaches, showing that it outperforms them substantially.
Finally, we use the LTR model to filter a dataset of posts that discuss the 189 conspiracy claims in our dataset, which we then use for further analysis.
In the following, we describe these steps in detail.

\descr{Building ground truth annotations.}
We first randomly sample 50 conspiracy claims from our existing 189 Snopes claims to use as a training set for the ranking model.
We refer to these claims as \initialtrainclaims~.
We then manually annotate the ground truth keywords for each conspiracy claim to use in training. 
For each claim, the ground truth is a set of keywords made up of terms from the conspiracy claims, which produces the most accurate set of results when queried against our Reddit dataset, optimizing for both relevance and size of results.
We only use the posts data for building the ground truth to reduce the human efforts required to judge the relevance of comments; these are traditionally longer than the posts and are not self-contained (comments might require additional efforts to understand the context of the discussion).

Our ground-truth selection process is iterative.
We start by initializing \basekeyword~, which are two words that are part of each claim, consisting of the subject and the object of a conspiracy claim.
We query the Reddit post data store with the \basekeyword~ and retrieve a set of results.
Intuitively, the \basekeyword~ usually has a very broad context and will return many irrelevant results as false negatives.
To evaluate the effectiveness of our keywords, we check randomly sampled 20 posts from the returned results and then adjust the \basekeyword~ by either adding new words from the claim or removing the words which are causing the false positives.
We repeat the process until we find the most relevant set of keywords for the individual conspiracy claims.
We repeat the whole process for each claims in \initialtrainclaims~.
At the end of the human annotation process, we end up with the ground truth keywords for each of the 50 conspiracy claims.
This ground truth is then used to learn the ranking function to identify the best set of keywords.
At the end of the process, the LTR algorithm can select the best keyword from a list of potential query terms.
The potential query terms are derived from the conspiracy claims and should not require any external context.
The learning model on LTR is optimized to select the ground truth we annotated as the best keyword.

\descr{Data pre-processing.}
We remove stopwords from the conspiracy claims while doing basic pre-processing, e.g., lowercasing text and removing punctuation marks.
We break down the cleaned-up conspiracy claims to n-grams of length four, which are our potential set of query terms.
We then query our Reddit data store, retrieving all the posts matching the query terms of each of the potential query terms.
The results returned for the potential query terms are then used to extract our learning to rank model features, as discussed in the upcoming section.

\descr{Feature Engineering.}
LTR requires us to generate a dataset consisting of a query set, relevance information, and feature values to learn the ranking.
The most widely used benchmark to build models based on LTR is the LETOR dataset~\cite{qin2010letor}, which contains query sets, learning features, and labeled rankings related to the 25 million pages \textit{GOV2} web page collection.
The dataset uses 46 different features related to Web documents retrieval.
We cannot directly utilize the LETOR dataset since our data store is composed of posts from microblogs (Reddit and Twitter), which are fundamentally different from Web pages.
Instead, we take inspiration from the LTR dataset and develop seven features that we use in our LTR experiments.
In the following, we briefly discuss these features.
First, we use the term \spanningsubset~  to describe the oldest 20\%, newest 20\%, and a 10\% sample from between the oldest and the latest posts results returned from the query, based on the timestamp attached with their posts.

From there, we derive  seven features:
\begin{enumerate}
\item
Count of the total number of hits (posts and comments) produced by the query.
\item
The median pairwise similarity score between the entries of \spanningsubset~.
\item
The median similarity score between the entries in \spanningsubset~ and the conspiracy claim.
\item
The mean pairwise similarity score between the entries of \spanningsubset~.
\item
The mean similarity score between the entries in \spanningsubset~ and the conspiracy claim.
\item
The median value of the TextRank scores~\cite{mihalcea2004textrank} of the query terms.
\item
The median score of the Term Frequency-Inverse Document Frequency scores~\cite{ramos2003using} of the query terms.
\end{enumerate}
The similarity score between the set of results, and between the query and the results are calculated using the Word Mover’s distance~\cite{kusner2015word}.

\descr{Training the LTR model.}
We use the RankLib project, part of the Lemur Toolkit~\cite{lemurtoolkit} which includes a family of learning to rank algorithms.
We use all eight algorithms implemented by Lemur (MART, RankNet, RankBoost, AdaRank, Coordinate Ascent, LambdaMART, ListNet, and Random Forests) for our experiments.
To evaluate our model, we use a rank-based evaluation metric commonly used in information retrieval settings: Mean Average Precision (MAP).
We choose MAP for evaluation because it is well suited for the binary nature of the rankings (a query keyword is either relevant to the conspiracy claim or not).

Due to the inherent nature of the keyword extraction problem, our dataset has many incorrect combinations (majority class) of keywords compared to one or two correct combinations (minority class) of keywords as the optimal query.
For this reason, we have to use an undersampling algorithm as this imbalance caused the learning algorithms to perform very poorly.
Undersampling techniques remove examples from the training dataset that belong to the majority class to reduce class distribution skew.
We experiment with multiple undersampling techniques available in the \textit{Imbalanced-learn} library~\cite{lemaitre2017imbalanced} such as the Near Miss Undersampling, Condensed Nearest Neighbor Rule Undersampling, Tomek Links Undersampling, Edited Nearest Neighbors Rule for Undersampling, among which the NearMiss-3 algorithm produced the best results.
After applying NearMiss-3 to filter out the representative samples for the false query set, each conspiracy theory claim has one or two combinations of ground truth keywords and the same number of queries with the False label. %

To test our LTR model, we perform 5-fold cross-validation on our ground truth.
Random Forest with LambdaMART~\cite{burges2010ranknet} as the bagging ranker produces the best results.
We further increase performance by performing a grid search over hyper-parameters (number of leaves, number of bags, number of trees, and minimum leaf support).
The 5-fold cross-validation on our ground truth using the tuned Random Forest model achieves a MAP of 0.75.
As we will show later in this section, this convincingly
outperforms other state-of-the-art keyword extraction approaches.
We further refer to this LTR model as \initialtrainmodel~.

\descr{Validating the LTR model.}
The previous experiment showed that our LTR approach could train an accurate model on our training dataset.
We now want to understand whether our model generalizes and can effectively identify posts related to claims that are not in the training set.
To this end, we design and conduct two experiments.

In the first experiment, we select 40 additional Snopes claims not part of the ground truth set, referring to them as \validationclaims~.
These claims were posted on Snopes after September 2020, while those from the ground-truth dataset were posted before September 2020.
The output keywords for these new claims can be inferred from the previously trained model. However, they will still require their respective ground truth label to evaluate the method.
Hence, we generate the ground truth for the \validationclaims~ through the same iterative method we used for \initialtrainclaims~.
Similarly, we generate the potential query terms set for the \validationclaims~, following the same steps for \initialtrainclaims~.
We then query the Reddit data store and generate the feature values for candidate keywords for each claim.
Finally, we perform inference on these new claims to see our trained model can identify the best set of keywords.
Our model achieves a MAP of 0.782, indicating our approach effectively identifies keywords and retrieves data for previously unseen claims.
Additionally, this experiment also demonstrates the efficacy of our model on a small window of discussion scenarios.
The model was validated on claims produced after September 2020, which is a comparatively smaller window of discussion of the conspiracies on social media than the discussion window of conspiracies the model was initially trained on (Pre-September 2020).
This effectiveness on small window data implies the model helps detect emerging conspiracy theory discussions early in their formation on social media.
Finally, since we have the ground truth of 40 additional Snopes claims from  \validationclaims~, we expand our training set to a total of 90 claims, which we call \expandedclaims~.
We retrain a new LTR  model on the \expandedclaims~, which we refer to further as \expandedmodel~.

In the second experiment, we aim to verify the learned model is not biased towards the social media platform it was trained on (i.e., Reddit) and can be applied
to cross-platform detection of conspiracy theories.
To this end, we draw 40 new claims that were not a part of training the \expandedmodel~ and call them \twittervalidationclaims~.
Following the same steps as for the \initialtrainclaims~ and \validationclaims~, we generate the potential query terms set for the \twittervalidationclaims~, query the Twitter dataset, and generate the feature values for each of the candidate keywords for the claim.
Note we previously generated the features from results queried on Reddit for validation, while we build the features from results queried on Twitter for this experiment.
We use the previously trained \expandedmodel~ model for inference on the \twittervalidationclaims~.
This experiment achieves a MAP of 0.794, showing that our LTR model is portable to other platforms.

\begin{figure}[t]
\centering
\includegraphics[width=0.8\linewidth]{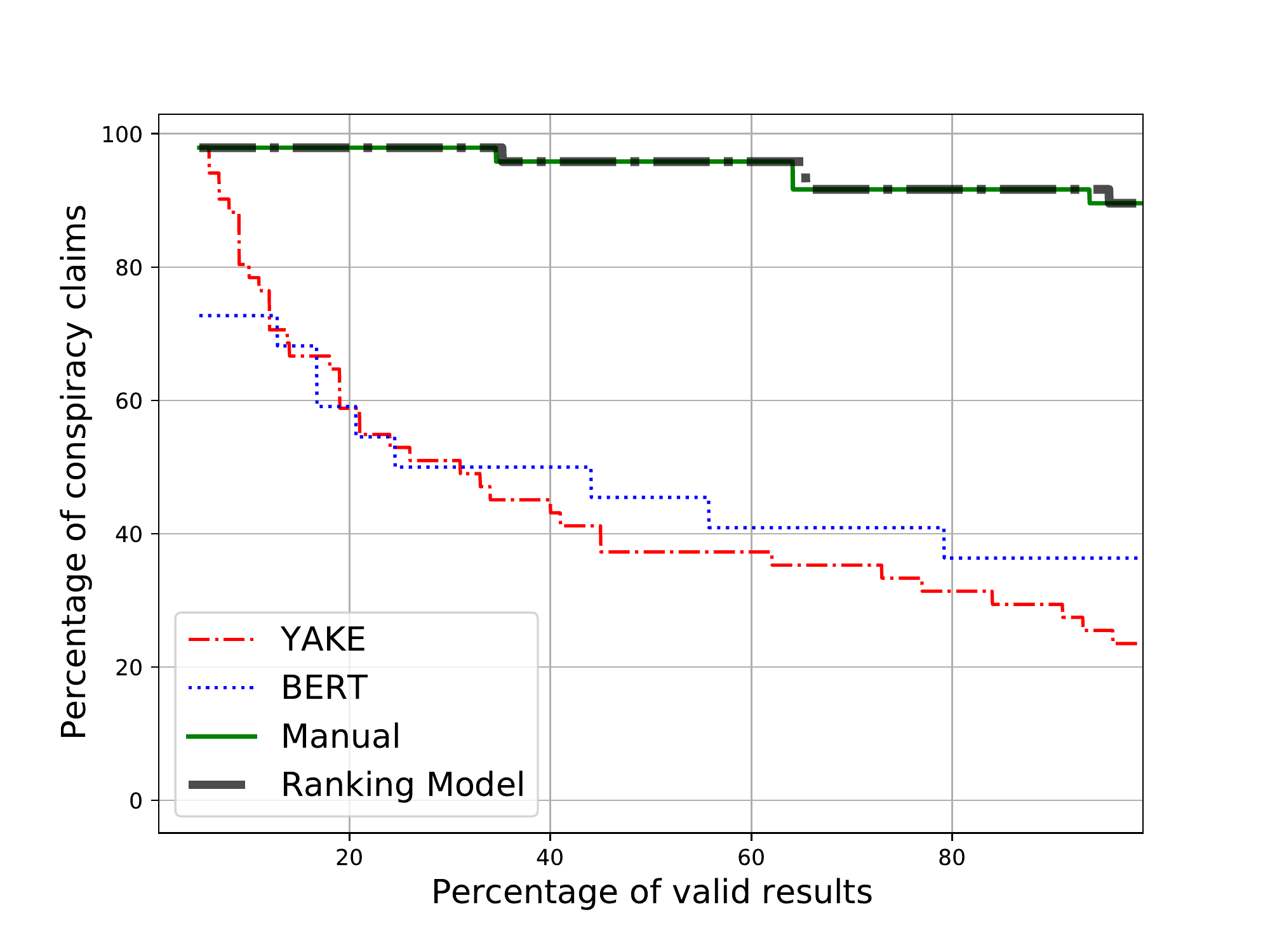}
  \caption{Validating the performance of various keyword identification methods.\vspace{-0.15in}}
\label{fig:keywordvalidation}
\end{figure}

\descr{Comparison with other keyword extraction methods.}
\label{section:comparison_model}
While we showed that our LTR model effectively identifies a set of keywords that allows us to retrieve accurate posts based on a conspiracy claim, other keyword extraction algorithms have been proposed by the research community to achieve similar tasks.
We, therefore, compare our approach with the two state-of-the-art keyword extraction algorithms, YAKE~\cite{campos2018yake}, and KeyBERT~\cite{grootendorst2020keybert}.

We compare the keyword extraction methods on \expandedclaims~, which consists of 90 conspiracy claims.
The results of our comparison are illustrated in Figure~\ref{fig:keywordvalidation} where we compare the percentage of valid results returned by the keywords returned from the different algorithms, including our LTR model.
Yake and KeyBert perform well only for a small number of claims, and the quality of their results drops dramatically the more claims they analyze.
On the other hand, our approach achieves a similar performance as the manual process that we followed to build ground truth, demonstrating that our technique is ideal for identifying conspiracy theory posts in the wild.

\descr{Extracting conspiracy posts from our dataset.}
After training our LTR model and validating its performance on ranking experiments with the other methods, we run it on the remaining 49 conspiracy claims to get the best set of keywords for the unannotated claims.
After applying the keywords, we obtain 66,724 total posts (33,285 from conspiracy-oriented subreddits and 33,439 from non-conspiracy-oriented subreddits) and 288,721 total comments (38,186 from conspiracy-oriented subreddits and 250,535 from non-conspiracy-oriented subreddits) from Reddit, and 379,708 tweets from Twitter.

Looking at the discussion on Twitter, the average number of tweets per conspiracy theory
is 994, the median is 124, and the standard deviation is 2968.
Similarly, the conspiracy-oriented subreddits have an average of 149 posts per conspiracy theory, with a median of 60 and a standard deviation of 220 posts.
The conspiracy-oriented subreddits have an average of 149.25 comments per conspiracy discussion, with the median being 34, and the standard deviation is 290.
In contrast, the non-conspiracy-oriented subreddits have an average of 830 comments per conspiracy claim, with a median value of 123 comments, and a standard deviation of 1945.
In contrast, the non-conspiracy-oriented subreddits have an average of 830 comments per conspiracy claim, with a median value of 123 comments and a standard deviation of 1945.

\subsection{Discussion of Conspiracies}
\label{section:description_language}
\descr{Language of conspiracy discussion between communities.}
After extracting the dataset of relevant conspiracy posts, we aim to study how the conspiracy theories are discussed on the different Web communities.
To this end, we study the similarity of the language used by different communities to discuss the same conspiracy theories.
We train a separate Word2Vec model~\cite{mikolov2013distributed} for each conspiracy claim for each Web community.
We use the skip-gram model to train word embeddings, shallow neural networks aiming to predict the context of a specific word.
To train them, we use the full corpus (both posts and comments) about a conspiracy claim for each Web community, and all tweets for Twitter.
That is, each conspiracy claim will have eight different trained word2vec models.

We extract the vector embedding of the identified query keywords and compare their cosine similarity.
We align the pairwise vector embeddings by using the Procrustes matrix alignment method~\cite{ross2004procrustes} to compare vector similarity of the same word across two different models.
Since there are multiple keywords as part of the query set associated with a conspiracy claim, we compute the average of the cosine similarities across all keywords identified by our LTR model.
The results of this experiment are discussed in Section~\ref{section:lang_discussion}.

\subsection{Influence Estimation between Communities}
\label{section:description_influence}
Next, we study the temporal nature of the conspiracy discussions on the Web communities and in particular, the influence multiple communities have on each other in this context.
First, we create a time series capturing the cascades of each conspiracy claim per Web community.
Then, we model the influence between Web communities using a statistical framework known as Hawkes Processes~\cite{hawkes1971spectra}.
Hawkes Processes can be used to quantify the influence of each Web community on the others to the discussion of conspiracy theories.

In this paper, we follow three steps to calculate influence between communities using Hawkes Processes:

\begin{itemize}
 \item For each claim, we extract the posts, comments, and tweets discussing it and build a time series for each community.
   We consider each community as a separate process.
 \item We fit a Hawkes model for each conspiracy claim by following the approach discussed in~,\cite{linderman2014discovering,linderman2015scalable} which uses Gibbs sampling to infer the parameters of models from the data.
The approach that we follow automatically samples the background rates and the shape of the impulse responses between the processes.
 \item After calculating the influence that each community has on the others for each claim, we aggregate this influence to study the normalized influence for each community.
 We focused on the normalized influence since the total number of conspiracy discussions happening in each community might be different; hence the normalized influence gives us an approximation of the ``efficiency'' of a community to influence other communities.
 This allows us to understand what communities are particularly influential in spreading different types of conspiracy theories. %
\end{itemize}

\section{General Characterization}

\begin{figure*}[!t]
\minipage[t]{0.32\textwidth}
 \includegraphics[width=\linewidth]{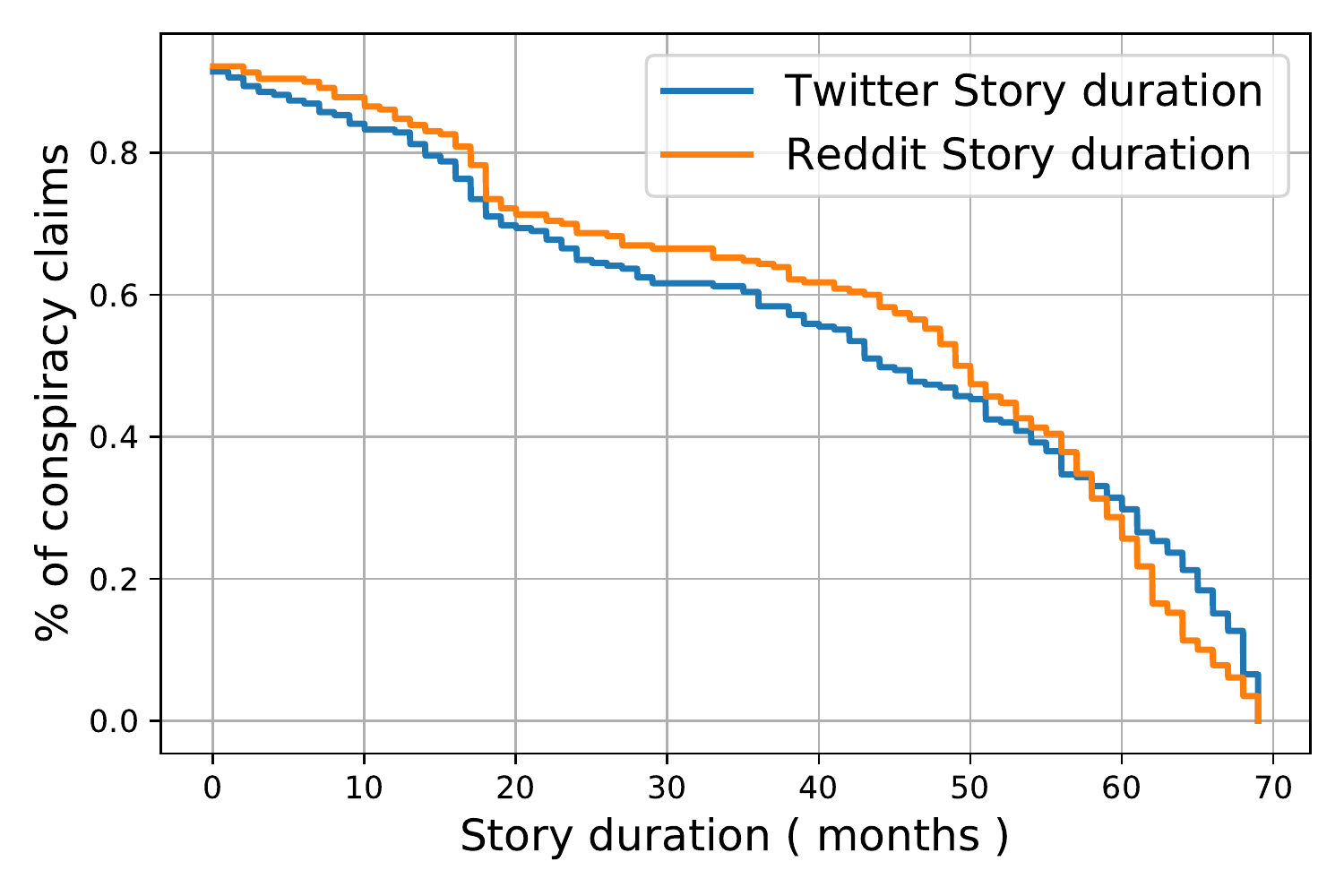}
 \caption{Duration of discussion of conspiracy on platforms.}\label{fig:storyduration}
\endminipage\hfill
\minipage[t]{0.32\textwidth}
 \includegraphics[width=\linewidth]{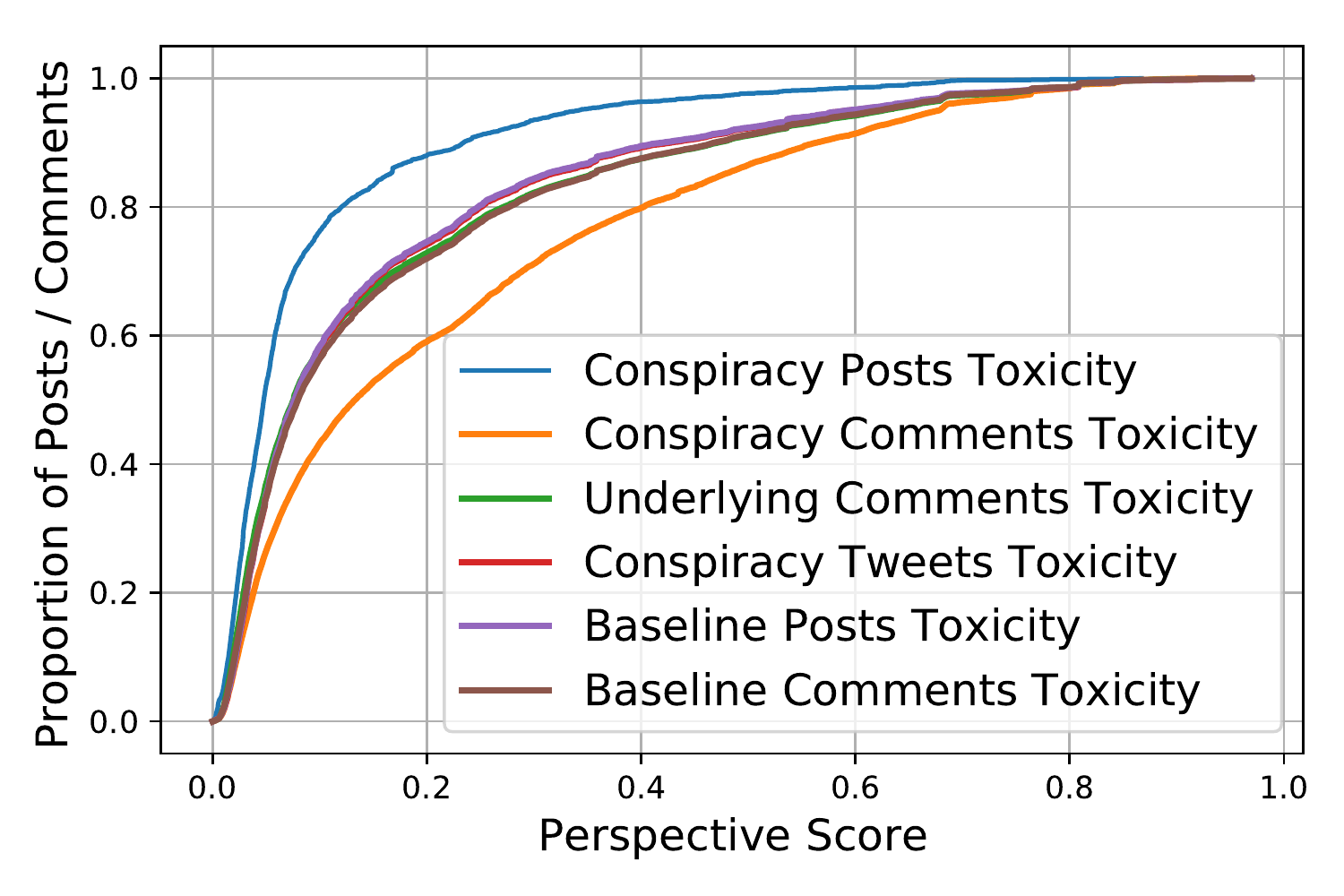}
 \caption{Distribution of Severe Toxicity score for conspiracy discussions.}\label{fig:toxicity}
\endminipage\hfill
\minipage[t]{0.32\textwidth}%
 \includegraphics[width=\linewidth]{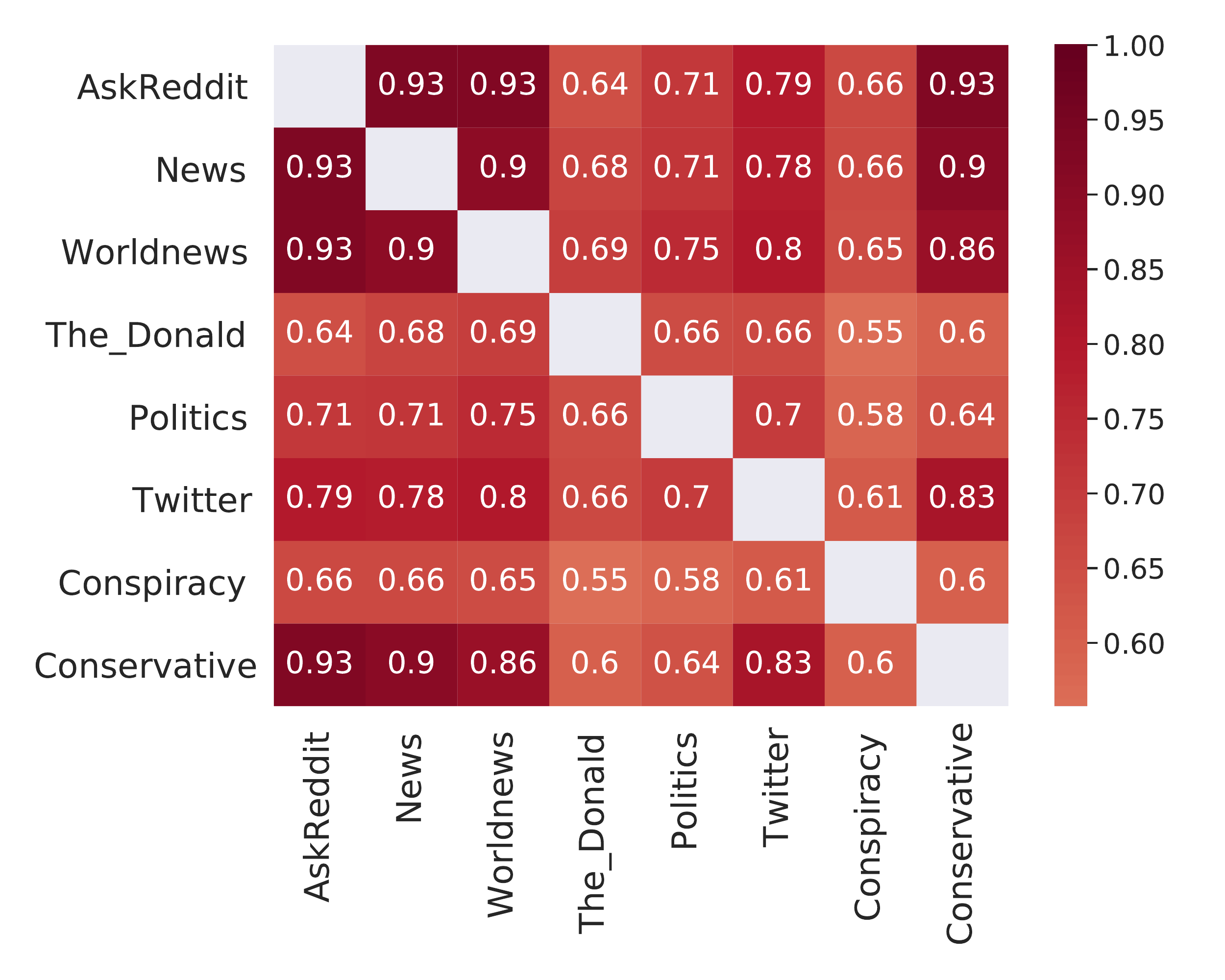}
 \caption{Discussion similarity between Web communities.}\label{fig:heatmap_similarity}
\endminipage
\end{figure*}

In this section, we first analyze how long conspiracy theories are discussed on social media.
We then look at the toxicity of language used to discuss them, comparing it to the general discussion on the same Web communities.

\descr{Lifespan.}
We define the lifespan of a conspiracy theory as the duration between the first appearance of a conspiracy discussion and the latest appearance on the respective Reddit and Twitter datasets.
In Figure~\ref{fig:storyduration}, we plot the Complementary Cumulative Distribution Function (CCDF) of the lifespan of conspiracy theories (in months) being discussed on both Reddit and Twitter.
Conspiracy theories have a similar lifespan on both Reddit and Twitter.
The vast majority of conspiracy theories are discussed for longer than a year (84.78\% on Reddit and 82.85\% on Twitter), with 29.79\% on Reddit and 25.65\% on Twitter being discussed for more than five years.

\descr{Toxicity of posts and comments.}
We use Google's Perspective API~\cite{perspectiveapi} to measure the toxicity of the discussion surrounding the conspiracy theories.
We use the {\em Severe Toxicity} model to characterize this aspect since previous work found it more reliable and less prone to ``false positives'' when measuring toxic speech~\cite{aliapoulios2021gospel,ribeiro2021evolution,ribeiro2021platform}.

For Reddit, we separate the posts and the comments discussing conspiracy theories. %
We also retrieve the comments replying to Reddit submissions of conspiracy theories to understand the toxicity of interaction of these discussions.
  We measure the Perspective score for these comments separately and refer to them as \textit{Conspiracy Submission Comments}.\footnote{These comments may or may not include the keywords related to the conspiracy claim from the submission.}

For Twitter, we look at the Severe Toxicity score for tweets that discussed conspiracy theories and were detected
from the method discussed in Section~\ref{section:learningtorank}.
To compare the toxicity measures against a baseline, we sample the same number of random posts and comments from the non-conspiracy-oriented subreddits discussed in Section~\ref{section:datacollection}.

  Figure~\ref{fig:toxicity} reports the cumulative distribution function (CDF) of the distribution of Severe Toxicity scores for our datasets.
We observe that the comments about conspiracy theories are the most toxic, and by a large margin.
Interestingly, while comments discussing conspiracy theories are more toxic than the general baseline, the same is not true for posts or tweets.
We can also observe that the baseline posts and comments are very close to each other in terms of the distribution of toxicity. In contrast, there is a substantial gap between the toxicity score of comments and posts.
The discussion on Twitter is also very close to the discussion on non-conspiracy-oriented subreddits but happens to be more toxic than the conspiracy posts themselves.
Finally, we observe that the  \textit{Conspiracy Submission Comments} happen to be more toxic than the posts under which the discussion happens but are far less toxic than the comments discussing conspiracy.

We assess the differences between these distributions by running a two-sample Kolmogorov-Smirnov (KS) test~\cite{kstest}.
We first compare the toxicity score in conspiracy posts and comments to the baseline set of posts and comments. We find that the differences between the following distributions are statistically significant at the $p<0.001$ level: conspiracy posts compared to the baseline posts (\textit{D}=0.207),  conspiracy comments compared to the baseline comments(\textit{D}=0.137).
As for conspiracy posts, and comments we have statistically significant differences between the distribution of conspiracy posts and comments ($p<0.001$, \textit{D}=0.337).
We also find that the toxicity scores of comments discussing conspiracy and conspiracy submission comments are statistically different ($p<0.001$, \textit{D}=0.151).
Finally, we find that the differences between the conspiracy posts and comments compared with the tweets discussing conspiracy are statistically significant ($p<0.001$), and so are conspiracy posts and tweets (\textit{D}=0.221), and conspiracy comments and tweets (\textit{D}=0.160).

\descr{Takeaways.}
In summary, we find that conspiracy theories on both Twitter and Reddit are discussed for long periods, with over 80\% of conspiracy theories on both platforms being discussed for over a year.
We find that Reddit comments discussing conspiracy theories are more toxic than general discussions.
While Reddit posts and tweets show closer toxicity to general posts, they still present statistically significant differences compared to the baseline.
\section{Language Analysis}
\label{section:lang_discussion}
As mentioned in Section~\ref{section:description_language}, we want to understand whether different communities discuss conspiracy theories differently.
For each claim, we take the word embedding model calculated for each community and compute the pairwise cosine similarity between all community pairs.
We then average these values across all claims and build a heatmap of the average similarity of conspiracy discussion in Web communities in Figure~\ref{fig:heatmap_similarity}.

We observe that, on average, conspiracy-oriented subreddits are the least similar to other communities with respect to language.
Discussion on Twitter maintains close similarity to the discussions across other non-conspiracy-oriented subreddits.
There are four subreddits (/r/AskReddit, /r/news, /r/worldnews, and /r/Conservative) which stand out as the most similar to each other with average cosine similarity higher than $0.9$.
The most similar remaining subreddits are /r/The\_Donald,  /r/politics, with an average similarity score of $0.66$ between them.
We also note that /r/AskReddit, which advertises itself as a ``place to ask and answer thought-provoking questions,'' is much closer to /r/news, /r/worldnews than the conspiracy-oriented communities.
Upon further investigation, some representative examples of conspiracy discussion occurring on /r/AskReddit are:
\begin{itemize}
\item ``can someone explain the supposed murder of vince foster by hillary clinton?''
\item ``trump's birther place ? donald trump was born in the country of pakistan and not in the united states  is it true ? how.''
\item ``is the child rape lawsuit against trump real or fake news?''
\end{itemize}
This implies that the discussion on /r/AskReddit is relatively shallow, mostly bringing up the news snippets from those news-based communities and asking if the claim is true or false.
This is in stark contrast to the more investigative and exploratory discussion of claims that happens in conspiracy-oriented subreddits. 

\descr{Takeaways.}
On Reddit, the discussion on /r/news, /r/worldnews, and /r/Conservative are closest to each other, while conspiracy-focused subreddits tend to show discussion that is further away.
This is also discussed in some selected case studies presented in the next section. %

\section{Influence Estimation}
\label{section:influence_estimation}

\begin{figure}[!t]
\centering
\includegraphics[width=0.98\linewidth]{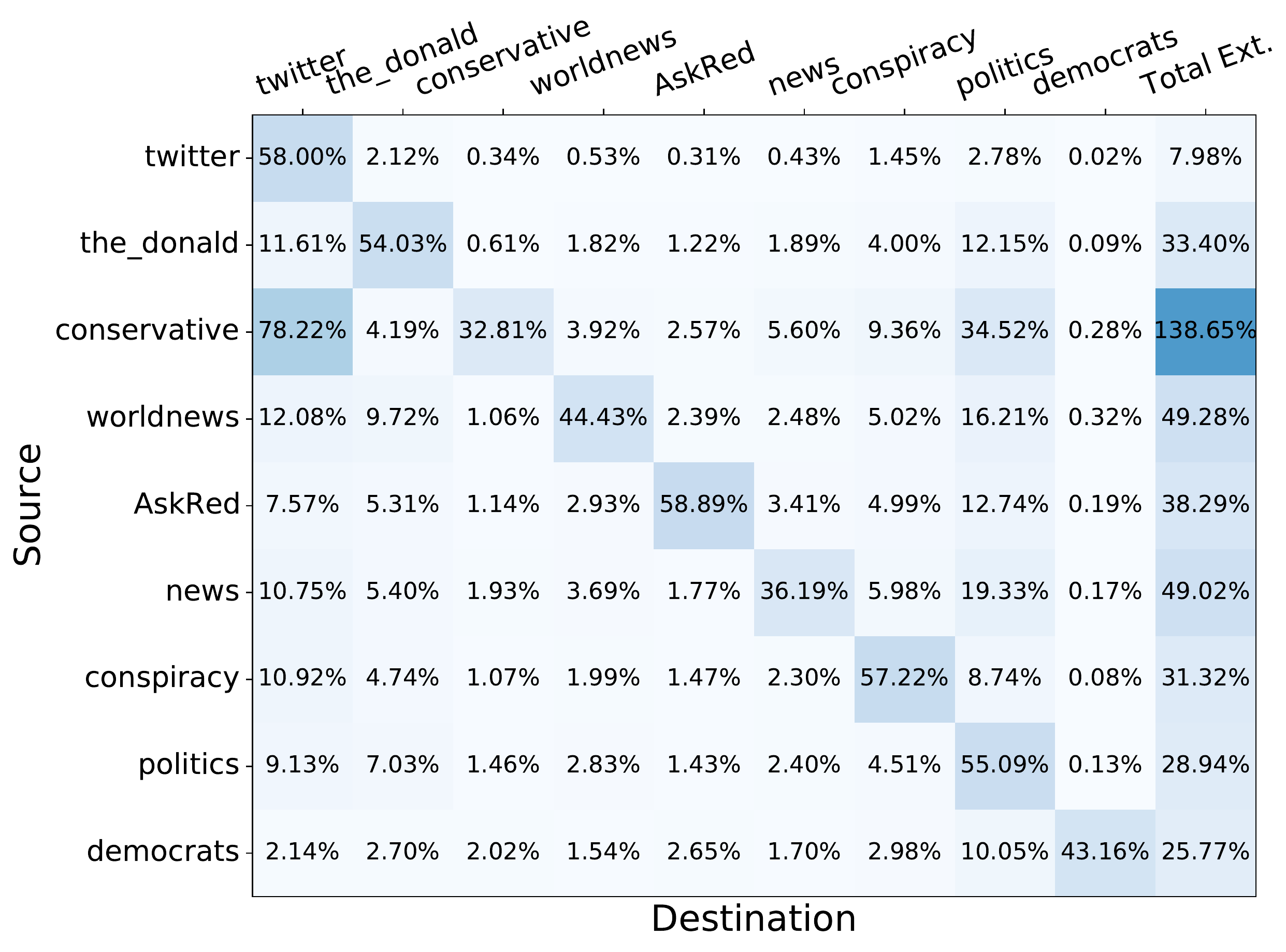}
\caption{Normalized Influence estimation: All conspiracies.}
\label{fig:influence_overall}
\end{figure}

We start by looking at the normalized influence\footnote{The total external influence is the sum of the normalized influence from a single source community to the rest of the platforms, and can amount to more than 100\%.} for each community for all the claims in our dataset; see Figure~\ref{fig:influence_overall}.
Note that /r/Conservative is by far the community with more influence on others, followed by /r/worldnews, /r/news, and /r/AskReddit.
In particular, /r/Conservative has the biggest influence on Twitter, indicating that the members of this subreddits are actively spreading conspiracy theories on that platform.
Conspiracy-oriented subreddits have a lower external influence, perhaps indicating that their users are mainly focused on dissecting conspiracy theories rather than spreading them to other communities.
Interestingly, Twitter has a much lower external influence than the other subreddits under study, which could be attributed to the much larger size of the platform compared to other communities.
The large size of the platform could explain the lower external influence of Twitter. 

Next, we are interested in investigating whether different communities are influential in spreading different types of conspiracy theories.
Based on our dataset, we select conspiracy claims that belong to three topics: Hillary Clinton (42 claims), Donald Trump (22 claims), and Covid-19 (44 claims).
Note that we do not include /r/The\_Donald community while computing the normalized influence for Covid-19 related claims because the subreddit was banned on June 29, 2020, which makes understanding Covid-19 related conspiracy on the community not possible.

\begin{figure*}[ht]
\minipage{0.325\textwidth}
 \includegraphics[width=\linewidth]{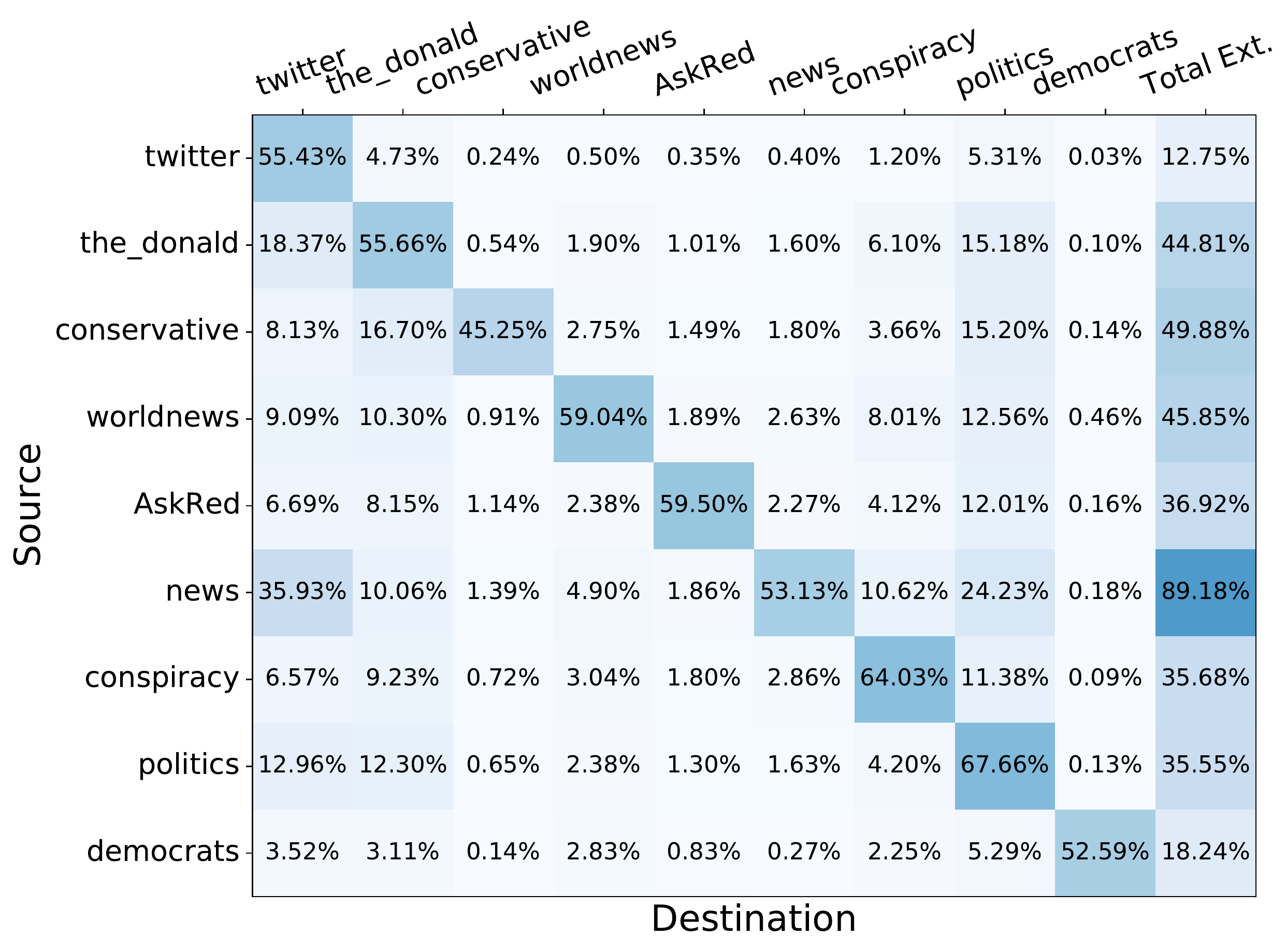}
 \caption{Normalized Influence estimation: Clinton related conspiracies.}\label{fig:influence_clinton}
\endminipage\hfill
\minipage{0.325\textwidth}
 \includegraphics[width=\linewidth]{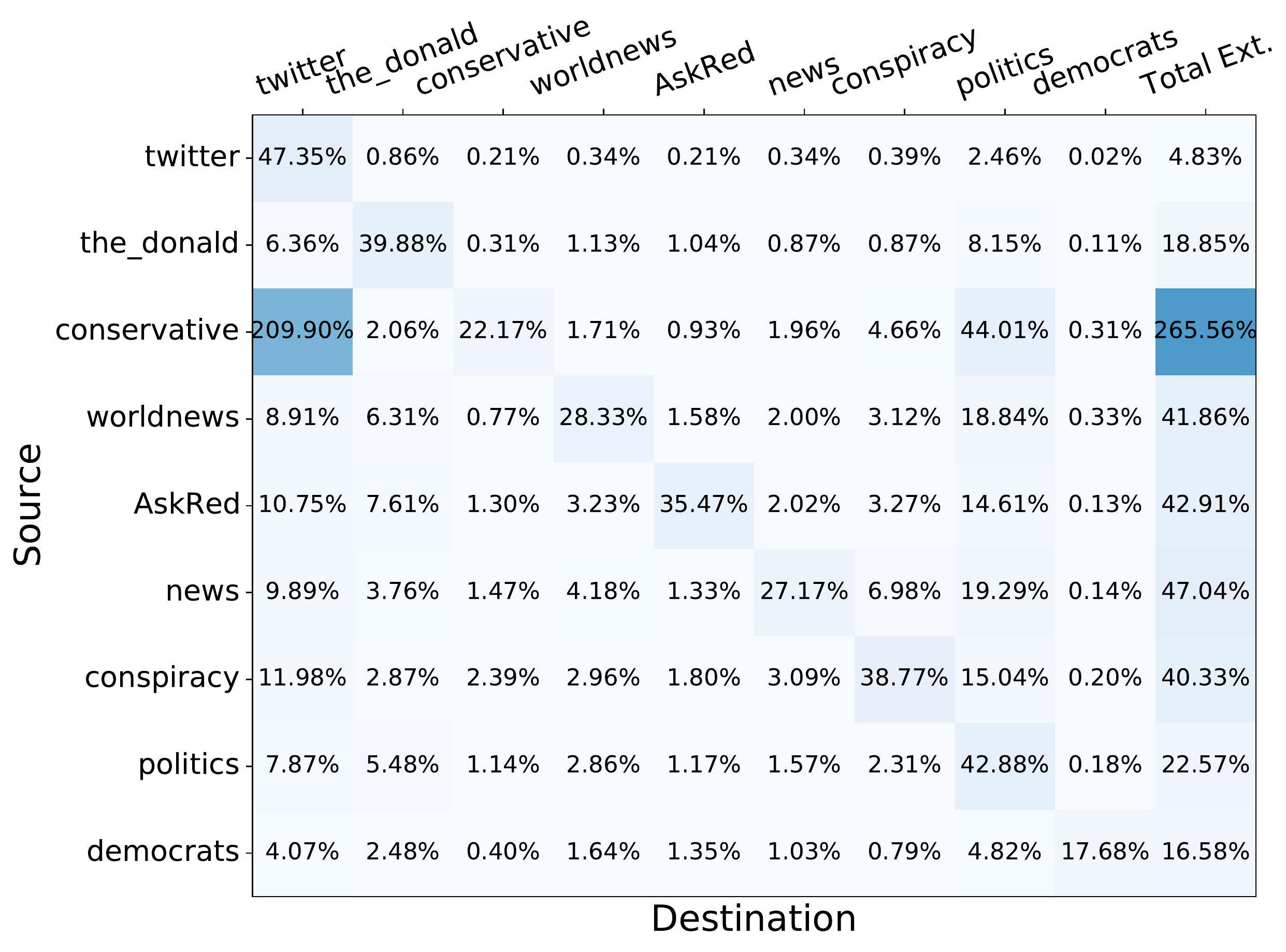}
 \caption{Normalized Influence estimation: Trump related conspiracies.}\label{fig:influence_trump}
\endminipage\hfill
\minipage{0.325\textwidth}%
 \includegraphics[width=\linewidth]{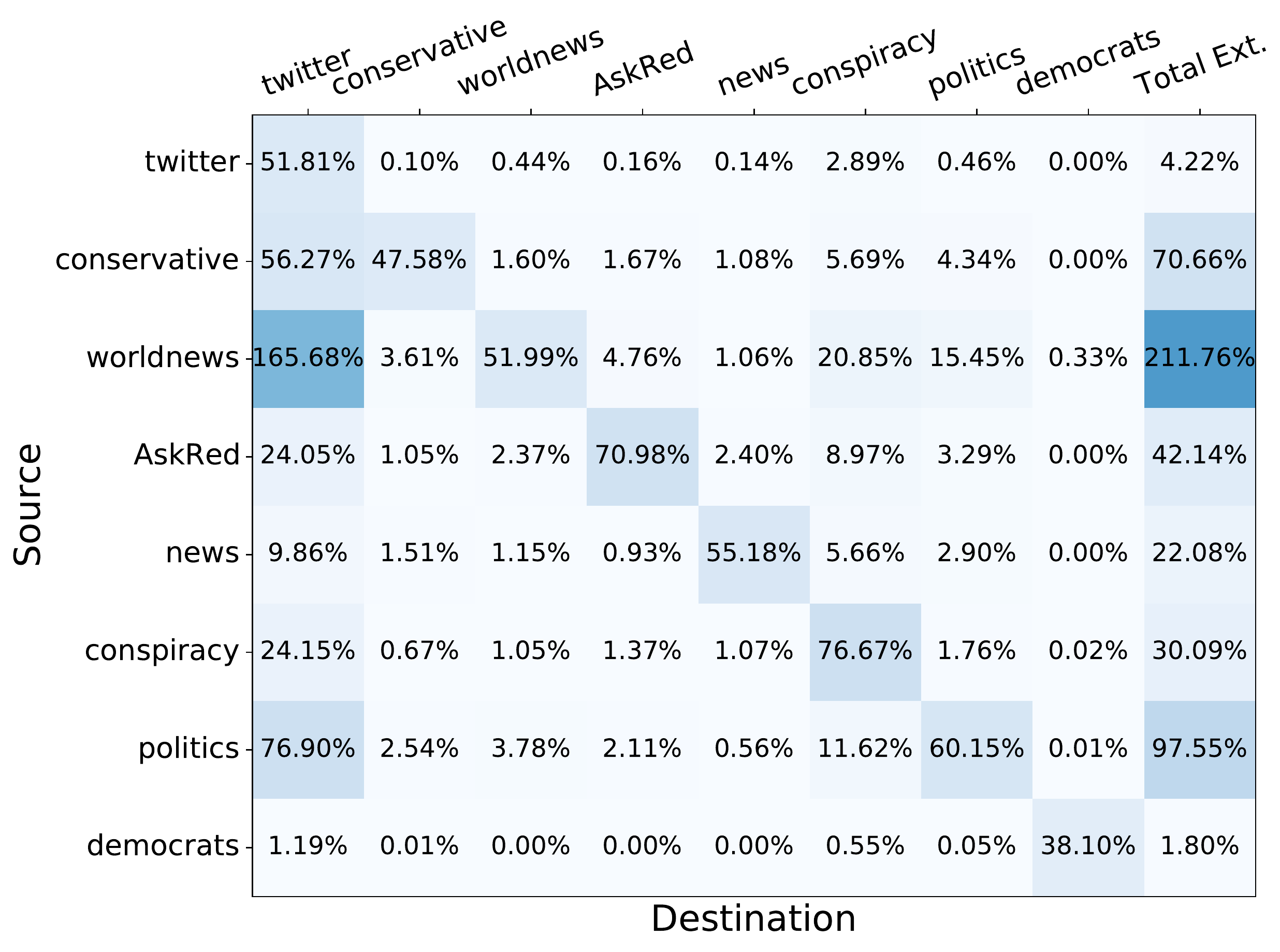}
 \caption{Normalized Influence estimation: Covid-19 related conspiracies.}\label{fig:influence_covid}
\endminipage
\end{figure*}

Figure~\ref{fig:influence_clinton} shows the normalized influence for each community regarding conspiracies related to Hillary Clinton.
/r/news has the most external influence for this type of conspiracy.
This is interesting because this subreddit is not dedicated to conspiracy discussion but rather to discussing news in general, yet, it is highly influential in spreading conspiracy theories externally to other platforms.
The second most influential Web community is /r/Conservative, which is perhaps expected since supporters of a political party have incentives in discrediting the candidate from the other party\jbnote{we can probably say something stronger about rcon, although not sure if we want to go down that road}.

Figure~\ref{fig:influence_trump} reports the normalized influence for each Web community regarding Donald Trump-related conspiracies.
Here we observe that the same trend as Clinton conspiracies does not hold: /r/Conservative is by far the most influential community in spreading Donald Trump-related conspiracies, with a particularly outsized influence on Twitter (209.9\%)
/r/Democrats, which could have an interest in discrediting Trump as a political opponent, actually have the second-lowest external influence after Twitter.
We observe that /r/The\_Donald is more influential in spreading Clinton-related conspiracies (44.81\%), and much less influential for Trump-related ones (18.85\%).
This might be indicative of the motivation of the community to actively smear and spread false rumors about Donald Trump's detractors/political opponents during the presidential election as well as a general avoidance of discussing conspiracies that paint in a bad light\jbnote{@PUJAN: PLEASE CHECK THIS SENTENCE TO MAKE SURE IT'S NOT LYING. THX!}. 

Finally, Figure~\ref{fig:influence_covid} illustrates the normalized influence of different Web communities with regards to Covid-19 related conspiracy theories; /r/worldnews is the most influential community, followed by /r/politics.
Both communities have an outsized influence on Twitter (165.68\% and 76.90\%, respectively).
This indicates that communities dedicated to general news discussion are fertile breeding grounds for the spread of conspiracies.

\descr{Case Studies: Qualitative Discourse Analysis}.
\label{section:qualitativediscourse}
While we find that /r/conservative, /r/worldnews, and /r/news are the most influential communities in spreading conspiracy theories, the actual discourse used in these communities may be different from the rest of the communities that we study, as the results in Section~\ref{section:lang_discussion} seem to suggest. 
To confirm this, we perform a qualitative analysis of two conspiracy theories for which /r/politics, /r/news, and /r/worldnews are the most influential.
The first conspiracy theory states that \emph{Rep. Marjorie Taylor Greene said that ``Jewish Lasers'' caused California wildfires}, while the second one states that \emph{Cesar Sayoc\footnote{A person who sent pipe bombs to Democratic officials in 2020.} was a lifelong Democrat who only recently covered his van in Trump Stickers}.
Our qualitative analysis finds that /r/news, /r/politics, and /r/worldnews merely report on the conspiracy theories without delving into the details. %
Looking at Twitter, we find that the platform users are mostly critical of the conspiracies and the actors spreading them. %
Conversely, more politically polarized communities (e.g., /r/The\_Donald) primarily discuss the conspiracy to defend the actors involved in it.
Finally, we find that the conspiracy-oriented subreddits discuss the conspiracies in greater detail, investigating the claim from a neutral and investigative angle.
Examples of comments from the various communities we studied can be found in the Appendix (omitted due to space reasons).

\descr{Takeaways.}
Our first finding is that non-conspiracy-oriented subreddits like /r/Conservative, /r/news, and /r/worldnews emerge as the most influential communities in spreading conspiracy theories and conspiracy dedicated communities are not as influential.
When we consider these results together with the findings from Section~\ref{section:lang_discussion}, they further reinforce that conspiracy-oriented subreddits mainly act as echo chambers when discussing conspiracy theories.

We also find that different communities are influential in spreading conspiracy theories on different topics.
While /r/news is the most influential subreddits for conspiracies related to Hillary Clinton, /r/conservative is the most influential for conspiracy theories related to Donald Trump.
Interestingly, /r/democrats is the least influential subreddit when discussing these conspiracies.
When looking at Covid-19 related conspiracies, /r/worldnews and /r/politics are the two most influential subreddits in spreading them.
This is quite alarming as it highlights that mainstream news communities play a major role in spreading health-related conspiracy theories.

Finally, a qualitative analysis of selected case studies shows that the detailed breakdown of the conspiracy theories happens away from the most influential communities, which are primarily limited to reporting related events.

\section{Discussion and Conclusion}

This paper presented a computational pipeline based on Learning-to-Rank to identify online posts discussing a set of known conspiracy theories. 
We used it to collect Twitter and Reddit discussion for a set of 189 known conspiracy theories identified by Snopes, finding hundreds of thousands Reddit submissions, comments, and tweets discussing them.
We make two main findings in how these conspiracies are discussed and spread on Web communities.

First, we find that there are quantitative differences in how conspiracy-oriented and more mainstream-oriented communities discuss conspiracy theories.
In short, conspiracy-oriented communities are very clearly involved in investigative and exploratory discussion; i.e., they are actively engaged in developing the theories, while more non-conspiracy-oriented communities primarily discuss conspiracy theories at a much higher level.
Second, we find that conspiracy-oriented communities themselves have relatively little influence concerning the overall spread of conspiracy theories.
Instead, the spread is mainly driven by more mainstream communities (e.g., /r/worldnews in the case of COVID-19 conspiracies).

In conjunction, these two findings have profound implications.
Our findings contribute to the growing body of evidence that simple solutions like banning/deplatforming worrying communities may not be as effective as they would seem at first glance~\cite{ali2021understanding,jhaver2021evaluating,ribeiro2021platform}.
For conspiracy theories, in particular, our results show that most of them are not driven by the communities with an evident devotion to them.
While conspiracy-oriented communities go deeper into the claims, they are relatively isolated with little influence outside their sphere.
This implies that the net effect of removing them would be minimal, at least in terms of quashing the prevalence of their discussion elsewhere.

This leads to the obvious question: what \emph{is} a good solution for this class of problems?
While we provide a data-driven method to \emph{detect} the discussion of conspiracy theories, alas, this does not say much about to \emph{do} about them.
That said, we do have promising suggestions for future work.
First, if hard moderation techniques are employed, our findings suggest that they should also target communities that use conspiratorial claims to push an agenda.
Although these are not actively involved in the creation of conspiracy theories, a good case can be made that they are \emph{weaponizing} them.
Moreover, early research results indicate that soft moderation techniques might help mitigate the spread of disinformation~\cite{sharevski2021misinformation,zannettou2021won}, and our method could be used to automate the use of these techniques.
For instance, it could be integrated into Reddit's automod bot to post a warning under conspiracy theory-related discussion.

\descr{Acknowledgments.}
This work was supported by the National Science Foundation under grants CNS-1942610, IIS-2046590, CNS-2114407, and CNS-2114411, a grant from the Media Analysis Ecosystems Group, and the UK's National Research Centre on Privacy, Harm Reduction, and Adversarial Influence Online (REPHRAIN, UKRI grant: EP/V011189/1).

{\small
\bibliographystyle{abbrv}
\bibliography{bibilography.bib}}

\appendix
\section{Case Studies}
\label{sec:appendix:case_studies}

\subsection{Case Study 1: Did Rep. Marjorie Taylor Greene Say ‘Jewish Lasers’ Caused California Wildfires?}

Most influential community: /r/politics, /r/news. 

\subsubsection{/r/politics, /r/news}
\label{snippet:jewishlaser_mainstream}
We first study how the subreddits which had the most influence in the discussion of the given conspiracy theory; /r/politics, and /r/news discuss the conspiracy.
We find that the discussion in these most influential subreddits is happening mostly on the line of bringing up the incident, and not going into the details of it trying to further discuss it.
A  few posts sampled from the community discussing the conspiracy theory follows:

\begin{tcolorbox}\small
\begin{compactitem}
\item marjorie taylor greene penned conspiracy theory that a laser beam from space started deadly 2018 california wildfire
\item marjorie taylor greene spread false theory that jews started camp fire with space laser
\end{compactitem}

\end{tcolorbox}

\subsubsection{Conspiracy-oriented Communities}
\label{snippet:jewishlaser_conspiracy}
While in the case of conspiracy-oriented communities, we observe that the discussions try to examine the conspiracy in depth; from both sides, about a neutral, investigative discourse. The discussion in this community attempts to examine the stupidity of the claim, as well as calls to examine it further with more introspection; despite how bizarre it might be.
A few posts sampled from the community discussing the conspiracy theory follows:
\begin{tcolorbox}\small
\begin{compactitem}
\item strange: it works for me   you can always just google it   the gist of it is:  ``jewish" space laser refers to a conspiracy theory about an in orbit space laser possibly created by rothschild inc  that is allegedly responsible for the california wildfires  the theory was notably purported by congresswoman marjorie taylor greene in a 2018 facebook post
\item What’s weird is i tried to look for the origin of her claims  she`s jumping to conclusions and making pretty wild connections but [never explicitly says anything about  jewish space lasers‚ (https://www newsweek com/marjorie taylor greene jewish space laser mockery 1565325) in her 2018 fb post. I'm  not trying to defend her: she's  nucking futs: but unless i'm missing something i don't see where the  jewish space laser meme came from
\item fun fact: not once did margorie taylor greene say anything about "jewish space lasers"  in her comment: she was wondering if people who claimed to see lasers starting the fire were actually seeing a mistargeted beam from a satellite made by a company called solaren ...
\end{compactitem}
\end{tcolorbox}

\subsubsection{Twitter}
We see the discussion regarding the conspiracy in Twitter play out interestingly, and differently than other platforms.
The discourse points towards being critical of the person who spread the conspiracy, and trying to hold them accountable if the actors of the conspiracy claim ( Marjorie Taylor Green in this instance) are present on the platform. 
They also invoke the conspiracy theory on other remotely relatable incidents, and due to the nature of the platform allowing them to do so; often confronting the actor directly about the conspiracy they spread.
A  few posts sampled from the community discussing the conspiracy theory follow:
\begin{tcolorbox}\small
\begin{compactitem}
\item I don't disagree. But taking advice from someone who didn't know about the holocaust and thinks Jewish space lasers start wildfires probably isn't great parenting.
\item Hey: remember that time you said a Jewish space laser started the CA wildfires?
\item Perhaps the woman who thinks wildfires are set by Jewish space lasers shouldn't be lecturing an epidemiologist on what measures are effective against the spread of an epidemic. Sit down: stay in your lane.
\end{compactitem}
\end{tcolorbox}

\subsection{Case Study 2: Was Cesar Sayoc a Lifelong Democrat Who “Recently” Covered His Van in Trump Stickers?}

Most influential community: /r/worldnews.
\subsubsection{/r/worldnews}

As we can see, the discourse nature in the community seems to be heavily focused regarding reporting of the news incidents, and the emerging details of the case.
The new details seem to be adding up to the common theme towards blaming Trump for the incident.
A  few posts sampled from the community discussing the conspiracy theory follows:
\begin{tcolorbox}\small
\label{snippet:sayoc_mainstream}
\begin{compactitem}

\item suspected maga bomber id'd as 'native american trump supporter' cesar sayoc.
\item pro trump mail bomb suspect cesar sayoc held without bail after new york court hearing
\item pipe bomb suspect cesar sayoc describes trump rallies as 'new found drug'
\item maga bomber cesar sayoc was radicalized by trump and fox news before terror plot: lawyer says

\end{compactitem}
\end{tcolorbox}

\subsubsection{/r/The\_Donald}
\label{snippet:sayoc_donald}
Analyzing how the discussion plays out in this community is particularly interesting as this subreddit is comprised primarily of Trump supporters.
We discovered that, as expected, the discussions were heavily communicating the idea that Trump shouldn’t be meddled in the whole incident, as he is not responsible for violence someone else started.
Strikingly different from other communities blaming Trump, the discussion in this community is trying to defend him, and criticize the general public reaction of how the mainstream media has falsely involved Trump on the whole story.
A few posts sampled from the community discussing the conspiracy theory follows:
\begin{tcolorbox}\small
\begin{compactitem}
\item just like james t  hodgkinson: the anti trump bernie fan who tried to massacre republican congressmen: apparent trump fan cesar sayoc is clearly mentally ill  bernie wasn't to blame for hodgkinson's actions: and trump isn't to blame for sayoc's actions
\item everything you wanted to know about cesar sayoc  yes: he appears to be a trump fan  if he's guilty: he did nothing violent; he's one bad apple out of how many millions of us?
\item clare lopez on twitter: more background on sayoc just doesn't add up   lots more investigation needed! cesar sayoc: maga bomber, facebook betrays democrat trump infiltrator: anti gop posts

\end{compactitem}
\end{tcolorbox}

\subsubsection{Conspiracy-oriented communities}
We find that discussion in the conspiracy-oriented communities regarding this conspiracy is very Trump critical, and discuss how his speech and actions are indirectly inciting violence.
A few posts sampled from the community discussing the conspiracy theory follows:
\begin{tcolorbox}\small
\label{snippet:sayoc_conspiracy}
\begin{compactitem}
\item i don't have cable  sorry man  it makes perfect sense that a bunch of paranoid: backwoods: gun loving morons who have been told that liberals and the press are the enemies of the people would do this  who have been told, 2nd amendment people   you know what to do:  requires no leap of faith whatsoever    edit: [aaaand we have a bingo](https://metro co uk/2018/10/26/suspected maga bomber identified native american trump supporter cesar sayoc 8079040/)
\item he preaches hate  he incites violence  he inspires attacks   we knew this before friday's  arrest of cesar sayoc: ....
\item viral photo does not show pipe bomb suspect cesar sayoc with a 'democrat donor' (but of course that does not stop crazy conspiracy theorists who seem to disregard at the same time the extensively documented pro trump activity of the magabomber)

\end{compactitem}
\end{tcolorbox}

\subsubsection{Twitter}
As observed in the previous example, we find that discussions on Twitter try to connect the instance of a discussion to a greater cause/issue.
They bring up the problem, be critical of Trump and his supporters, and recollect how activities such as in this incident have been a repeating pattern of violence over the years.
This line of a discourse of connecting the individual incidents related to the conspiracy theory to a larger problem/ context occurs to be a common theme in Twitter discussions.
A  few posts sampled from the community discussing the conspiracy theory follows:
\begin{tcolorbox}\small
\begin{compactitem}
\item RT : Rabid MAGA savages have been dangerous for years. Remember MAGA Bomber Cesar Sayoc: who sent pipe bombs to Trump's percieved enemies.
\item It seems like ages ago: but we had a warning how dangerous these lunatic Trump ppl were by this guy: Cesar Sayoc. Remember this asshole? He sent pipe bombs to Clinton: Obama: Deniro and CNN. Storming the Capitol isn't far-fetched for the cultists if you think about it.
\item RT : Violence inspired by Trump: Charlottesville, Tree of Life Synagogue, Capital Gazette, Cesar Sayoc pipe bombs, El Paso shootings.
\item Not to mention Cesar Sayoc and the Tree of Life Synagogue.  Those who stuck with Trump have a lot of blood on their hands.
\end{compactitem}
\end{tcolorbox}

\end{document}